\documentclass[cameraready]{Interspeech}

\title{VoiceChat-TTS: A Low-Latency Continuous Speech Synthesis Model for Interactive Agents}

\usepackage{multirow}
\usepackage{booktabs}

\makeatletter
\newcommand*\concat
  {\mathbin{\mathmakebox[\widthof{${+}\m@th$}]{+\hskip-1emplus1fil+}}}

\makeatother

 \newcommand\citep{\cite}

\usepackage{soulutf8}
\usepackage{todonotes}
\usepackage{placeins}
\usepackage{tikz}
\usepackage{svg}

\author[affiliation={1}, equalcontribution, correspondingauthor]{Edresson}{Casanova}
\author[affiliation={1}, equalcontribution]{
Jaehyeon}{Kim}
\author[affiliation={1}]{Mariana}{Graterol Fuenmayor}
\author[affiliation={1}]{Shehzeen}{Hussain}
\author[affiliation={1}]{Viacheslav}{Klimkov}
\author[affiliation={1}]{Valentin}{Mendelev}
\author[affiliation={1}]{Mikyas}{Desta}
\author[affiliation={1}]{Paarth}{Neekhara}
\author[affiliation={1}]{Piotr}{Zelasko}
\author[affiliation={1}]{Chen}{Chen}
\author[affiliation={1}]{Elena}{Rastorgueva}
\author[affiliation={1}]{Ke}{Hu}
\author[affiliation={1}]{Ankita}{Pasad}
\author[affiliation={1}]{Xuesong}{Yang}
\author[affiliation={1}]{Aya}{Alja'fari}
\author[affiliation={1}]{Rajarshi}{Roy}
\author[affiliation={1}]{Rohan}{Badlani}
\author[affiliation={1}]{Jason}{Roche}
\author[affiliation={1}]{Jason}{Li}
\author[affiliation={1}]{Zhehuai}{Chen}

\address{
    {NVIDIA Corporation}
  }

\email{ecasanova@nvidia.com}

\usepackage{comment}

\begin{document}

\maketitle

\begin{abstract}
    {
Spoken dialogue is a natural form of human--computer interaction, yet most speech language models remain limited to turn-based operation and lack real-time adaptability, such as user barge-in. Recent duplex speech-to-speech and speech-to-text models reduce latency by replacing multi-stage pipelines, but often compromise speech quality because accurate ASR, interruption handling, and high-fidelity synthesis must be optimized jointly. We propose VoiceChat-TTS, a low-latency, continuous, and streamable text-to-speech model for interactive agents. VoiceChat-TTS is driven directly by LLM text-token streams, supports explicit interruption via control tokens, and produces silence when no textual input is available. The model enables always-on, responsive speech generation while preserving modularity and high speech quality, and it supports mid-utterance interruptions without resetting the KV cache.
    }
\end{abstract}

\section{Introduction}\label{sec:intro}
Spoken dialogue is a natural and intuitive modality for human--computer interaction. However, most existing speech language models remain constrained to turn-based operation and lack real-time adaptability, such as support for user barge-in. Recent duplex speech-to-speech (S2S) \cite{hu2025salm,defossez2024moshi,roy2026personaplex} and speech-to-text \cite{lu2025duplexmamba} models achieve low latency and simplified deployment by replacing traditional multi-stage pipelines that rely on separate automatic speech recognition (ASR), voice activity detection (VAD), and text-to-speech (TTS) components. ASR systems transcribe spoken audio into text, VAD detects the presence or absence of speech to guide downstream processing, and TTS generates waveform output from text input.

While end-to-end duplex models are highly promising, they often exhibit degraded speech synthesis quality. This degradation largely arises from the substantial data and modeling challenges involved in jointly optimizing accurate ASR, robust interruption handling, and high-quality speech generation within a single architecture.

Prior work on speech-to-speech (S2S) models has explored architectures that incorporate dedicated speech decoders connected to a shared backbone through latent representations \cite{demo-s2s-asru,xu2025qwen3}. Although effective in reducing overall interaction latency, these approaches increase architectural complexity and can compromise the modularity and debuggability of traditional pipeline systems. 

Recently, streaming TTS has made significant progress in reducing time-to-first-audio (TTFA) and enabling incremental speech generation. One early example is the streaming speech decoder of Audio Flamingo 3-Chat~\cite{goel2025audio}, which was designed for low-latency conversational speech generation by consuming streamable text inputs and producing speech tokens progressively. This design demonstrated that high-quality neural speech synthesis can be integrated into interactive speech-language systems without requiring the complete text response in advance. Subsequent streaming TTS systems further improved latency and alignment. For example, VoXtream~\cite{torgashov2025voxtream} uses an incremental decoder-only Transformer with monotonic alignment and limited look-ahead to map incoming phonemes to acoustic tokens, while SpeakStream~\cite{bai2025speakstream} employs a decoder-only architecture trained with interleaved text-and-speech sequences to absorb streaming text from large language models (LLMs). In addition, Qwen3-TTS~\cite{hu2026qwen3} explores efficient text-to-speech alignment for real-time applications by directly ingesting LLM tokens to achieve very low TTFA while maintaining high synthesis quality and expressiveness.

However, these systems primarily address streaming generation within a single response. They do not natively model the continuous, always-on behavior required by full-duplex interactive agents, where the speech decoder must remain active across conversational time, generate silence when no agent text is available, and stop promptly in response to user barge-in. VoiceChat-TTS builds on this line of streaming speech decoders, particularly the Audio Flamingo 3-Chat speech decoder, and extends it with explicit silence modeling, interruption control tokens, and a unified training formulation for both single-turn and multi-turn conversational synthesis.



In this paper, we propose VoiceChat-TTS, a continuous, streamable, and low-latency text-to-speech model designed for interactive agents. VoiceChat-TTS is driven directly by large language model (LLM) text-token streams, supports explicit interruption through control tokens, and produces silence when no textual input is available. This design enables always-on, responsive speech generation while preserving modularity and high synthesis quality. Furthermore, VoiceChat-TTS is compatible with both duplex interaction frameworks and speech-to-speech pipeline systems, and avoids resetting the KV cache when generation is interrupted mid-utterance.

The main contributions of our work are as follows:

\begin{itemize}
 \vspace{-0.05cm}
    \item We introduce VoiceChat-TTS, a continuous, streamable, and low-latency TTS model that directly consumes LLM text-token streams and generates silence when no agent text is available;
    \item We propose a reliable control-token-based interruption mechanism that halts ongoing speech and transitions the output to silence during mid-utterance user barge-ins;
    \item We present a unified training strategy that combines high-quality single-turn TTS data with complex multi-turn conversational data while minimizing the distribution mismatch between the two settings;
    \item We demonstrate that VoiceChat-TTS achieves competitive speech quality relative to strong offline and streaming baselines while meeting the latency and interruption-handling requirements of interactive agents.
\end{itemize}

VoiceChat-TTS code is publicly available in NVIDIA NeMo Speech\footnote{https://github.com/NVIDIA-NeMo/Speech}, and the model checkpoint is available on Hugging Face as part of NVIDIA-NemotronLabs-VoiceChat-11B\footnote{https://huggingface.co/nvidia/NVIDIA-NemotronLabs-VoiceChat-11B}.

\section{VoiceChat-TTS Model}
\label{sec:duplex-eartts}
\begin{figure*}[]
\centering
\resizebox{0.9\textwidth}{!}{%
\includegraphics[width=1\textwidth]{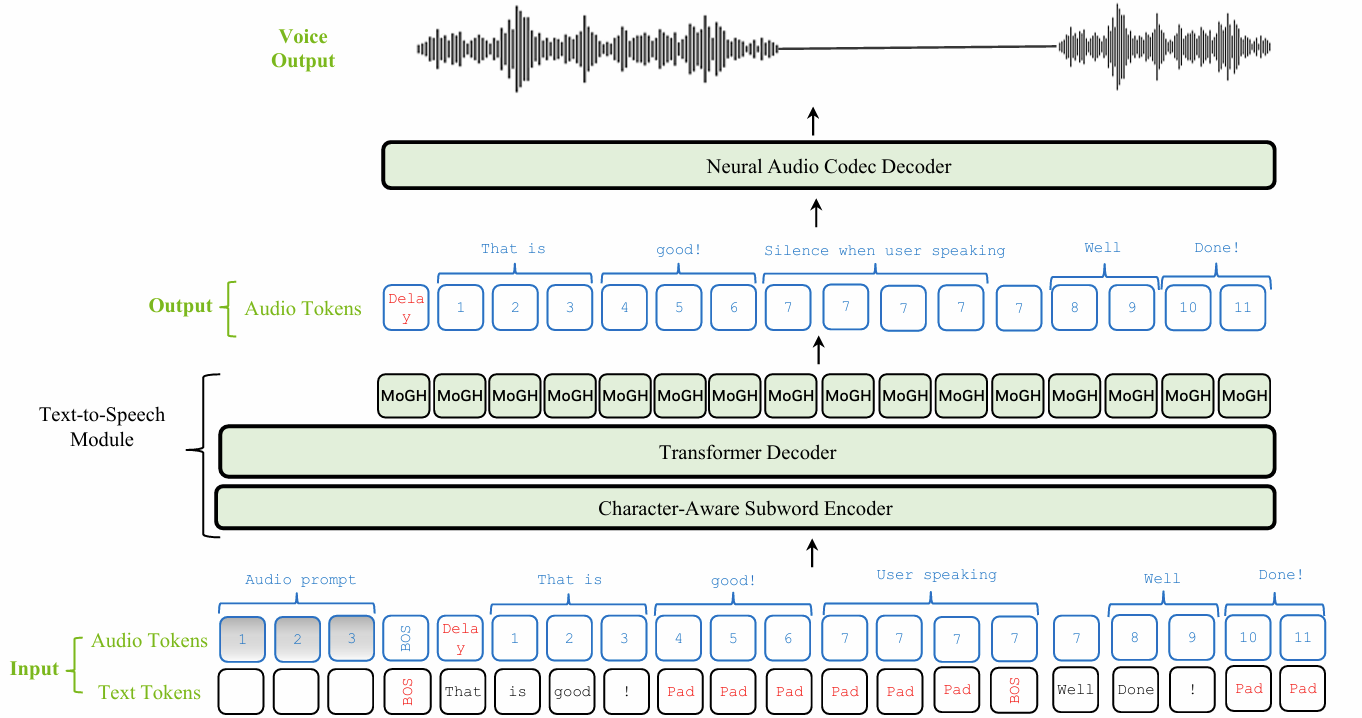}
}
 \vspace{-0.3cm}
 \caption{VoiceChat-TTS architecture overview.}
  \vspace{-0.3cm}
 \label{fig:arch}
\end{figure*}

The proposed architecture builds directly upon the streaming speech decoder of Audio Flamingo 3-Chat~\cite{goel2025audio}, incorporating several modifications to support full-duplex interactions. The Audio Flamingo 3-Chat streaming speech decoder was designed for low-latency conversational use cases and real-time interactions. It relies on streamable inputs, allowing text to be processed incrementally without requiring the full sequence in advance, and produces streamable outputs by generating speech tokens progressively for immediate audio playback. However, fully duplex interactions impose additional operational requirements: the system must handle mid-sentence user interruptions and produce silence while the user is speaking to support natural turn-taking. To fulfill these requirements, we implement several modifications on top of the base Audio Flamingo 3-Chat streaming speech decoder architecture. Figure~\ref{fig:arch} shows an overview of the VoiceChat-TTS architecture. 

\textbf{Audio Codec:} The audio codec backbone is a fully causal autoencoder composed entirely of convolutional layers, similar to that of~\cite{goel2025audio}. In our work, we train the codec to compress 22~kHz waveforms at a frame rate of 12.5~Hz using 31-codebook Residual Vector Quantization (RVQ) tokens. In this configuration, each generated audio-token frame corresponds to an 80~ms waveform chunk. This frame rate is compatible with recent duplex speech-to-speech models such as~\cite{hu2025salm,defossez2024moshi}. For efficient streaming codec inference, the network caches the inputs of the convolutional blocks only within their receptive field.

\textbf{Text Tokenizer:} In contrast to Audio Flamingo 3-Chat~\cite{goel2025audio}, we use the NVIDIA Nemotron Nano 2 subword tokenizer~\cite{nvidia2025nvidianemotronnano2} and augment it with Beginning-of-Sequence (BOS) and interruption tokens. On the LibriTTS \textit{test-clean} subset, the tokenizer produces text tokens at an average rate of 4.16~Hz, corresponding to approximately one text token for every three acoustic-token frames. During training, the text stream is right-padded to match the length of the acoustic sequence, enabling incremental processing throughout the interaction. A BOS token marks the beginning of each assistant turn, while an interruption token marks the point at which the model should stop speaking and transition to silence. We additionally introduce a one-token delay between the text and audio channels, with the text stream leading the audio stream. This provides limited linguistic look-ahead, ensuring that the model observes at least two subword tokens before generating the corresponding speech.

\begin{figure}[]
\centering
\resizebox{0.5\textwidth}{!}{%
\includegraphics[width=1\textwidth]{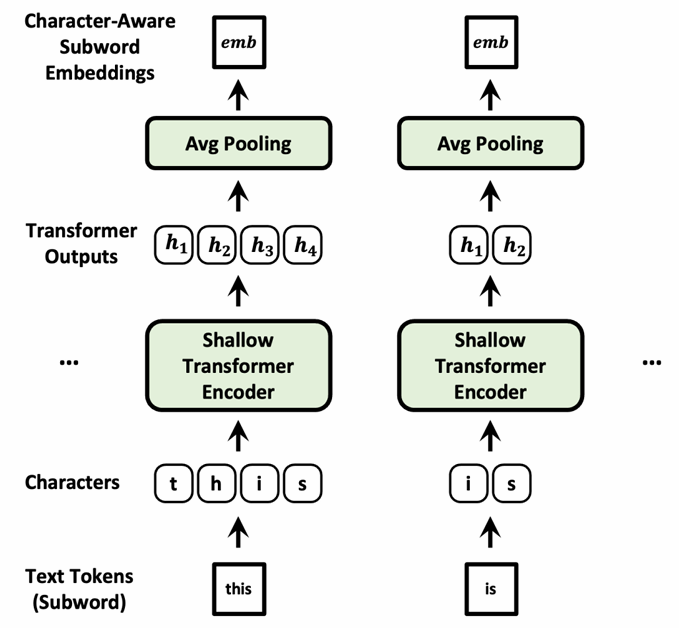}
}
 \vspace{-0.3cm}
 \caption{Character-Aware Subword Encoder architecture overview.}
  \vspace{-0.3cm}
 \label{fig:CAS}
\end{figure}

\textbf{Character-Aware Subword Encoder:} A prevalent issue in incremental text-to-speech models that consume LLM subword tokens is the mismatch between LLM vocabularies and TTS training corpora; many subwords produced by the former are rare or absent in the latter. To mitigate this problem, we introduce a Character-Aware Subword Encoder. Each input subword is first converted into a sequence of characters, which is processed by a shallow Transformer encoder. We then average-pool the character-level outputs to obtain robust character-aware embeddings that can generalize to unseen subwords. Figure~\ref{fig:CAS} shows an overview of the Character-Aware Subword Encoder. In addition, to improve pronunciation accuracy during incremental generation of long or fragmented words, we incorporate a dedicated embedding for subword-continuation tokens. This allows the model to process and synthesize multi-token linguistic units more cohesively.

\textbf{Mixture of Gaussian Estimation Head:} To accelerate generation of the deep RVQ token hierarchy, we integrate the Mixture of Gaussian head (MoGH)~\cite{kim2025efficient,goel2025audio} into the architecture. Instead of relying on a 31-step autoregressive decoding pipeline, this head iteratively unmasks the RVQ tokens. In each iteration, it predicts the continuous embedding vector of the masked RVQ tokens using MoG estimation and subsequently quantizes the vector into progressively unmasked discrete tokens. This iterative refinement is controllable, and prior results indicate that 4 to 8 iterations are sufficient to achieve high-fidelity acoustic reconstruction~\cite{goel2025audio}.

\textbf{Audio Prompt Conditioning:} The Audio Flamingo 3-Chat streaming speech decoder does not employ explicit speaker-reference conditioning. Instead, it is trained on concatenated utterances from the same speaker and learns to infer speaker identity from preceding acoustic context. This approach can be less effective at the beginning of generation, when little or no speaker-informative speech is available, and in conversational settings, where the recent acoustic context may be dominated by silence while the user is speaking. To provide an explicit speaker cue, we condition VoiceChat-TTS on a 3-second reference audio prompt. During training, the corresponding audio tokens prefill the beginning of the acoustic sequence, and the loss over this prompt region is masked so that the model uses the prompt as conditioning context rather than learning to reconstruct it. During inference, the same 3-second prompt initializes the acoustic context before synthesis begins.

\textbf{Boundary Embeddings and Gated Fusion:} Since our training objective targets multi-turn conversations, we introduce learnable text embeddings for the BOS and interruption tokens. These embeddings provide explicit boundary signals that improve the smoothness of conversational turn-taking. We also address an architectural instability in the base decoder, where high-magnitude RVQ embeddings can cause mixed-precision overflow and destabilize early-layer activations. We mitigate this issue by applying a gated fusion mechanism between incoming text embeddings and acoustic speech embeddings, ensuring stable numerical scaling throughout the network.

The final VoiceChat-TTS model comprises 977M parameters in total, including a 778M-parameter Gemma 3-based streaming TTS module~\cite{gemmateam2025gemma3technicalreport} and a 199M-parameter codec model.


\section{Experiments}
\label{sec:exps}

To train the VoiceChat-TTS model, we utilize a combination of standard single-turn text-to-speech corpora, synthetically generated multi-turn dialogues, and real-world conversational datasets. This diverse data mixture is essential for ensuring robust acoustic quality while simultaneously teaching the model complex interaction dynamics, such as turn-taking and interruption.

\subsection{Single-turn data}
To establish a robust foundation for acoustic generation, we leverage large-scale, standard text-to-speech corpora for the single-turn training phase. Specifically, we use approximately 70,159 hours of English speech sourced from the \emph{train-clean-360} and \emph{train-clean-100} subsets of LibriTTS~\cite{zen2019libritts}, the original HiFiTTS corpus~\cite{bakhturina21_interspeech}, an extended 36.7k-hour subset of HiFiTTS-2~\cite{langman2025hifitts}, and a proprietary 62-hour dataset comprising two high-fidelity speakers. We additionally incorporate a 32k-hour internal dataset derived from high-quality, publicly available YouTube videos under the Creative Commons license. 

A critical challenge in training duplex models is the distribution shift between single-turn generation, which typically begins speech synthesis immediately, and multi-turn interactions, which often begin with extended periods of silence while the user speaks. To mitigate this modality discrepancy, we apply a targeted data augmentation strategy to 50\% of our single-turn data. We randomly prepend between 0.5 and 5.0 seconds of background silence to the beginning of the audio sequences, shifting the corresponding text alignments accordingly. This technique effectively simulates a multi-turn dialogue context where the agent is initially in a listening state. By bridging this gap, we prevent the model from trivially distinguishing between standard TTS tasks and duplex scenarios.

\subsection{Multi-turn data}

\textbf{Synthetic datasets:}
To explicitly model complex duplex behaviors, we develop a curated synthetic multi-turn dataset comprising approximately 2.5k hours of speech. Millions of textual dialogue scripts are generated by open-source LLMs using NeMo Data Designer, with prompts tuned to cover a wide range of topics and situations. Each dialogue contains up to 15 turns and includes LLM-generated paralinguistic annotations. We then synthesize the scripts using a pipeline based on Chatterbox~\cite{chatterbox2025} and Koel-TTS~\cite{hussain2025koel}, producing multi-speaker dialogues with speaker identities drawn from the LibriTTS training sets. Crucially, 50\% of this synthetic data is designed to simulate overlapping speech and user barge-ins. This provides the alignment and explicit control tokens needed for the model to learn interruption handling. Furthermore, to expand phonetic coverage and improve robustness to out-of-vocabulary terms in conversational settings, we synthesize specialized multi-turn interactions containing complex words and proper names. In these interactions, each turn consists of four distinct words, systematically increasing the lexical diversity observed during duplex training. 

\textbf{Real conversation datasets:}
While synthetic data provides precise control over interaction timing and interruption labels, real conversational data is important for capturing natural dialogue dynamics. To this end, we incorporate unscripted data from the Fisher corpus alongside a small internal conversational dataset. These data allow the model to learn nuanced acoustic transitions, spontaneous backchannels, and natural turn-taking patterns that are difficult to fully reproduce through synthetic generation alone. Accurately transcribing paralinguistic events in spontaneous speech remains challenging; consequently, these datasets often contain untranscribed acoustic phenomena, such as laughter, sighs, and subtle backchannels, which the model must handle robustly.


\subsection{Experimental setup}
We train VoiceChat-TTS in two stages. First, we pretrain the model for 1.4M steps using only single-turn TTS data on 8 NVIDIA A100 GPUs. This stage provides a strong acoustic and linguistic foundation before introducing more complex conversational behavior. We then fine-tune the model for 200k steps on 32 NVIDIA H100 GPUs using the full training mixture, including single-turn TTS data, word and proper-name lists, real conversational speech, and synthetic duplex dialogues.

During fine-tuning, we use Lhotse~\cite{zelasko2021lhotse} dynamic bucketing to construct batches containing approximately 240 seconds of audio. The training mixture consists of 40\% single-turn TTS data, 15\% word and proper-name list data, 10\% real conversational data, and 35\% synthetic duplex conversational data. To prevent any modality discrepancy during training, single-turn and multi-turn samples are processed identically. In both formats, text tokens are right-padded with special padding IDs within the text channel until they match the total temporal length of the current conversational turn \cite{hu2025salm,demo-s2s-asru}.  

For all training stages, we use the AdamW optimizer with a learning rate of $4 \times 10^{-5}$.

\subsection{Results and Discussion}
We evaluate VoiceChat-TTS in terms of intelligibility, speaker similarity, and predicted overall speech quality, following the evaluation protocol of~\cite{hussain2025koel}. Intelligibility is measured using ASR-based character error rate (CER) and word error rate (WER), where transcriptions are produced by Parakeet-TDT~\cite{xu2023efficient}. Speaker similarity is measured as the cosine similarity between speaker embeddings extracted from the synthesized speech and the reference prompt audio using \textit{TitaNet-Large}~\cite{koluguri2022titanet}. We denote this metric as speaker embedding cosine similarity (SECS). Overall speech quality is estimated using Squim-MOS~\cite{kumar2023torchaudio}; we use it as a non-intrusive quality estimate rather than as a direct measure of naturalness.

At inference time, we use multinomial top-$p$ sampling with $p=0.95$, a classifier-free guidance (CFG) scale of $0.2$, and a noise scale of $0.001$ during Mixture-of-Gaussians (MoG) sampling. We use $8$ MoGH refinement iterations for all quality and latency evaluations. Because generation is stochastic, each experiment is repeated ten times and we report mean values with 95\% confidence intervals. Fixed reference outputs are evaluated once.

For unseen-speaker evaluation, we use the full \textit{test-clean} subset of LibriTTS~\cite{zen2019libritts}. For seen-speaker evaluation, we use five speakers observed during training. To evaluate the model in multi-turn settings, we construct synthetic multi-turn test conversations by grouping LibriTTS utterances from the same speaker. This setup allows us to measure whether synthesis quality remains stable as the number of generated turns increases.

\setlength{\tabcolsep}{3pt}
\begin{table*}[t]
\vspace{-2mm}
\caption{{Comparison between TTS models. VoiceChat-TTS is evaluated with different numbers of consecutive turns.}}
\centering
\resizebox{\textwidth}{!}{%
\begin{tabular}{llc|cccc|cccc}
\toprule
\multicolumn{3}{c}{} & \multicolumn{4}{c}{\emph{Unseen Speakers}} & \multicolumn{4}{c}{\emph{Seen Speakers}} \\
\toprule
Model & Type & \# Turns 
& CER(\%) $\downarrow$ 
& WER(\%) $\downarrow$ 
& SECS $\uparrow$ 
& Squim-MOS $\uparrow$ 
& CER(\%) $\downarrow$ 
& WER(\%) $\downarrow$ 
& SECS $\uparrow$ 
& Squim-MOS $\uparrow$ \\
\midrule
Ground Truth & - & - 
& $0.48$ 
& $1.40 $ 
& $0.830$ 
& $4.457$ 
& - & - & - & - \\
\midrule
Chatterbox-TTS~\cite{seo2026chatterbox} & Offline & - 
& $0.45 \pm 0.02$ 
& $1.24 \pm 0.02$ 
& $\mathbf{0.887 \pm 0.001}$ 
& $4.27 \pm 0.003$ 
& - & - & - & - \\
Audio Flamingo 3-Chat~\cite{goel2025audio} & Streaming & - 
& $2.85 \pm 0.33$ 
& $4.51 \pm 0.41$ 
& $0.761 \pm 0.002$ 
& $3.60 \pm 0.009$ 
& - & - & - & - \\
Qwen3-TTS-12Hz-1.7B~\cite{hu2026qwen3} & Streaming & - 
& $\mathbf{0.34 \pm 0.02}$ 
& $\mathbf{1.01 \pm 0.05}$ 
& $0.827 \pm 0.001$ 
& $\mathbf{4.45 \pm 0.006}$ 
& - & - & - & - \\ 

\midrule
VoiceChat-TTS & Streaming & 1 
& $1.00 \pm 0.10$ 
& $2.00 \pm 0.10$ 
& $0.757 \pm 0.004$ 
& $4.380 \pm 0.004$ 
& $1.20 \pm 0.10$ 
& $2.40 \pm 0.10$ 
& $0.785 \pm 0.001$ 
& $4.365 \pm 0.001$ \\
VoiceChat-TTS & Streaming & 2 
& $1.00 \pm 0.10$ 
& $1.90 \pm 0.10$ 
& $0.710 \pm 0.001$ 
& $4.377 \pm 0.002$ 
& $0.80 \pm 0.10$ 
& $1.90 \pm 0.10$ 
& $0.789 \pm 0.005$ 
& $4.358 \pm 0.004$ \\
VoiceChat-TTS & Streaming & 3 
& $1.10 \pm 0.10$ 
& $2.10 \pm 0.10$ 
& $0.696 \pm 0.001$ 
& $4.381 \pm 0.003$ 
& $0.80 \pm 0.10$ 
& $1.80 \pm 0.10$ 
& $0.778 \pm 0.001$ 
& $4.362 \pm 0.001$ \\
VoiceChat-TTS & Streaming & 4 
& $1.20 \pm 0.20$ 
& $2.20 \pm 0.20$ 
& $0.685 \pm 0.005$ 
& $4.376 \pm 0.003$ 
& $0.80 \pm 0.10$ 
& $1.80 \pm 0.10$ 
& $0.778 \pm 0.001$ 
& $4.359 \pm 0.001$ \\
\bottomrule
\end{tabular} 
}
\label{tab:tts_comparison_multi-turn}
\vspace{4mm}
\end{table*}

Table~\ref{tab:tts_comparison_multi-turn} compares VoiceChat-TTS with strong offline and streaming TTS baselines. Conventional TTS systems, such as Chatterbox-TTS and Qwen3-TTS, achieve the best single-turn intelligibility, speaker similarity, and predicted overall speech quality. However, these models are optimized for standard single-response synthesis and are not designed for continuous duplex interaction, where the decoder must remain active across conversational time, generate silence when no agent text is available, and respond reliably to interruption control signals.

We also include the Audio Flamingo 3-Chat streaming speech decoder~\cite{goel2025audio}, the closest architectural baseline to VoiceChat-TTS. Compared with this baseline, VoiceChat-TTS substantially improves intelligibility and predicted overall speech quality. In the one-turn setting, VoiceChat-TTS reduces WER from $4.51\%$ to $2.00\%$ and increases Squim-MOS from $3.60$ to $4.38$, while achieving comparable speaker similarity. 
These results show that the complete VoiceChat-TTS system improves upon the Audio Flamingo 3-Chat streaming decoder while extending it to continuous multi-turn generation, silence modeling, and interruption-aware synthesis.

VoiceChat-TTS achieves competitive predicted overall speech quality while supporting the additional requirements of interactive agents. In the unseen-speaker setting, the model obtains a Squim-MOS of approximately $4.38$ across all evaluated turn counts, close to the strongest baselines. Intelligibility also remains stable as the number of turns increases from one to four, with WER varying only from $2.00\%$ to $2.20\%$. For seen speakers, all metrics remain similarly stable across multi-turn generation: CER and WER remain low, Squim-MOS stays around $4.36$, and SECS remains within a narrow range from $0.778$ to $0.789$. This small seen-speaker similarity gap indicates that VoiceChat-TTS can preserve speaker identity consistently across multiple turns when the target speaker is well represented during training.

For unseen speakers, SECS decreases more noticeably as the number of generated turns increases, from $0.757$ for one turn to $0.685$ for four turns. This suggests that maintaining speaker identity over extended continuous generation remains more challenging in zero-shot or unseen-speaker settings, particularly when the model alternates between speech and silence states. Improving long-context speaker consistency for unseen speakers is therefore an important direction for future work, potentially through stronger prompt conditioning or explicit speaker-consistency objectives.


Finally, we compare VoiceChat-TTS with the PersonaPlex speech-to-speech (S2S) model~\cite{roy2026personaplex}. PersonaPlex jointly predicts text and audio tokens. For this comparison, we use its outputs on the Smooth Turn Taking subset of Full-Duplex-Bench~\cite{lin2025full}. To isolate the contribution of the speech decoder, we keep the PersonaPlex-predicted text-token stream fixed and resynthesize it with VoiceChat-TTS. We preserve the original temporal alignment of this stream, including PAD tokens during intervals in which the user is speaking. VoiceChat-TTS therefore runs continuously over the complete conversation timeline and is expected to generate silence during these intervals. CER and WER are computed from the full, untrimmed assistant output; consequently, any intelligible speech generated during an intended silence interval contributes ASR insertion errors. This setup evaluates both speech generation quality and the model's ability to remain silent during user speech while preserving the same linguistic content and timing. Since speaker identities are not matched between systems, we omit speaker-similarity metrics from this comparison.

\begin{table}[t]
\vspace{-2mm}
\caption{{Comparison between PersonaPlex and VoiceChat-TTS on the Smooth Turn Taking subset of Full-Duplex-Bench, using the same PersonaPlex-predicted textual content.}}
\label{tab:personaplex_voicechattts}
\centering
\begingroup
\setlength{\tabcolsep}{4pt}
\renewcommand{\arraystretch}{1.15}
\resizebox{\columnwidth}{!}{%
\begin{tabular}{l|ccc}
\toprule
Model 
& CER(\%) $\downarrow$ 
& WER(\%) $\downarrow$ 
& Squim-MOS $\uparrow$ \\
\midrule
PersonaPlex~\cite{roy2026personaplex} 
& $4.06$ 
& $5.00$ 
& $4.094$ \\

\begin{tabular}[c]{@{}l@{}}
PersonaPlex\\
+ VoiceChat-TTS
\end{tabular}
& $\mathbf{2.05 \pm 0.78}$ 
& $\mathbf{2.42 \pm 0.80}$ 
& $\mathbf{4.292 \pm 0.007}$ \\
\bottomrule
\end{tabular}
}
\endgroup
\vspace{-2mm}
\end{table}

Table~\ref{tab:personaplex_voicechattts} shows that resynthesizing the PersonaPlex-predicted text with VoiceChat-TTS improves both intelligibility and predicted overall speech quality while preserving the original linguistic content and text-token timing. VoiceChat-TTS reduces CER from $4.06\%$ to $2.05\%$ and WER from $5.00\%$ to $2.42\%$. Because these metrics are computed on the full, untrimmed assistant waveform, any intelligible speech produced during PAD-designated user-speech intervals contributes ASR insertion errors. The lower error rates therefore indicate both more intelligible synthesis of the intended content and limited intelligible speech leakage during intervals in which the model is expected to remain silent. VoiceChat-TTS also improves Squim-MOS from $4.094$ to $4.292$, indicating higher predicted perceptual quality. Overall, these results demonstrate the benefit of using VoiceChat-TTS as a modular, high-quality speech decoder in continuous S2S pipelines.

\subsection{Interruption Evaluation}
\label{sec:interruption}

To evaluate whether VoiceChat-TTS responds correctly to explicit interruption signals, we construct an FDB-timed controlled-interruption benchmark based on the User Interruption subset of Full-Duplex-Bench (FDB). This evaluation should not be interpreted as an official FDB score. We use FDB only for the user audio and interruption-timing annotations; because the benchmark does not provide canonical assistant transcripts suitable for TTS-only evaluation, we generate controlled assistant responses with known reference text. Following the Full-Duplex-Bench v1.5 stop-latency protocol~\cite{lin2025full}, we insert the interruption token at the detected onset of the interrupting user speech and measure how rapidly the active assistant output transitions to silence.

Speech activity is detected independently in the original FDB user audio and the assistant-only output generated by VoiceChat-TTS. Both waveforms are resampled to 16~kHz and processed using the official FDB \texttt{evaluation/get\_timing.py} implementation, which uses Silero VAD. Following the official configuration, adjacent user and assistant speech segments separated by less than 0.6~s and 0.5~s, respectively, are merged. We identify the interrupting user segment as the merged user segment with the greatest temporal overlap with the interruption interval provided in the FDB metadata. The onset of this segment defines the interruption time. Silero VAD detects active assistant speech at interruption onset in all 200 controlled examples.

We report three timing-based metrics. {Interruption Obey Rate at 320~ms (IOR@320ms)}, is the percentage of examples in which the assistant speech segment active at interruption onset ends within 320~ms. {FDB-v1.5-compatible Stop Latency} is computed from the user--assistant overlap intervals returned by the official \texttt{latency\_stop\_list}; the table reports the mean after pooling these intervals across examples. Finally, {Leakage@1s} is the cumulative duration of detected assistant speech during the first second after interruption onset, averaged across examples.

We additionally report {After-Interruption Character Error Rate (AI-CER)} to evaluate whether the decoder can recover from an interruption and synthesize the subsequent assistant turn without resetting its KV cache. For each example, we isolate only the assistant-output segment corresponding to the controlled turn following the interruption, beginning at its known BOS timestamp and ending at the turn boundary. We transcribe this segment, compute CER against the corresponding reference text, and average the resulting CER values across examples. Audio between interruption onset and the subsequent BOS—including any residual overlap, leakage, or silence—is excluded. AI-CER therefore measures the intelligibility and content preservation of the subsequent assistant turn, complementing the stop-latency and leakage metrics rather than incorporating the interrupted portion of the output. The evaluation contains all 200 examples from the FDB User Interruption subset, and every example is processed for each reported metric.

We compare two inference settings. When \emph{Force Silence} is false, the model must transition to silence solely through its learned response to the interruption token. When \emph{Force Silence} is true, inference injects a fixed silence audio-token frame at positions aligned with the interruption token, deterministically steering the generated speech stream toward silence. To obtain this frame, we encode a prolonged segment of pure silence with the codec and select the most frequently occurring 31-token codec frame.

\begin{table}[t]
\vspace{-2mm}
\caption{Effect of deterministic silence forcing on the FDB-timed controlled-interruption benchmark. Timing and leakage metrics are computed using the FDB v1.5 Silero VAD pipeline. AI-CER is computed only on the subsequent assistant turn and excludes the interruption-overlap interval. Values are means over ten stochastic runs; 95\% confidence intervals are omitted for compactness.}
\label{tab:voicechat_interruption}
\centering
\setlength{\tabcolsep}{3pt}
\resizebox{\columnwidth}{!}{%
\begin{tabular}{lcccc}
\toprule
\shortstack{Force\\Silence}
& \shortstack{IOR@320ms\\(\%) $\uparrow$}
& \shortstack{Mean stop\\latency (ms) $\downarrow$}
& \shortstack{Leakage@1s\\(ms) $\downarrow$}
& \shortstack{AI-CER\\(\%) $\downarrow$} \\
\midrule
False
& $96.8$
& $228.3$
& $169.1$
& $0.255$ \\
True
& $\mathbf{100.0}$
& $\mathbf{89.9}$
& $\mathbf{55.8}$
& $\mathbf{0.095}$ \\
\bottomrule
\end{tabular}
}
\vspace{-2mm}
\end{table}

Table~\ref{tab:voicechat_interruption} shows that VoiceChat-TTS learns a strong response to the interruption token even without deterministic enforcement, achieving an IOR@320ms of 96.8\%. Enabling \emph{Force Silence} increases IOR@320ms to 100.0\%, reduces mean FDB-v1.5-compatible Stop Latency from 228.3~ms to 89.9~ms, and reduces Leakage@1s from 169.1~ms to 55.8~ms. 
AI-CER also decreases from 0.255\% to 0.095\%. Because AI-CER excludes all audio preceding the subsequent BOS, this result indicates that deterministic silence forcing does not impair the model's ability to recover and generate the next assistant turn without resetting its KV cache. Overall, the results show that the learned interruption behavior is already reliable, while deterministic silence forcing provides stricter interruption compliance and substantially reduces residual speech after interruption.

\subsection{Latency}
\label{sec:latency}

We evaluate streaming latency using inter-token latency (ITL), defined as the wall-clock time required to produce one frame of acoustic tokens after receiving a streaming text update. VoiceChat-TTS operates at 12.5~Hz, so each acoustic-token frame corresponds to 80~ms of audio. Therefore, an ITL below 80~ms indicates that the acoustic-token predictor runs faster than real time.

For comparison, we use Qwen3-TTS-12Hz, which also operates at 12.5~Hz and is designed for low-latency streaming synthesis~\cite{hu2026qwen3}. The Qwen3-TTS technical report provides latency numbers for its 12~Hz models, including first-packet latency, tokenizer decoding time, and steady-state LM time per packet. However, those measurements are reported using the authors' internal vLLM engine on a ``single typical computational resource'', without specifying a directly comparable GPU configuration. To obtain a more controlled comparison, we remeasure Qwen3-TTS-12Hz on the same RTX A6000 setup used for VoiceChat-TTS and report latency under the same measurement protocol. Both models are optimized using the vLLM-Omni framework.

The final VoiceChat-TTS model comprises 977M parameters in total, including a 778M-parameter Gemma 3-based streaming TTS module~\cite{gemmateam2025gemma3technicalreport} and a 199M-parameter codec model. In contrast, the Qwen3-TTS-12Hz model used in this comparison is the 1.7B variant~\cite{hu2026qwen3}. Thus, VoiceChat-TTS is smaller while targeting the same 12.5~Hz streaming regime.

Table~\ref{tab:latency_itl} reports latency at concurrency levels 1 and 4. We separately measure acoustic-token ITL and codec decoding time. Acoustic-token ITL measures the time required by the neural decoder to produce the next frame of acoustic tokens. Codec latency measures the time required to decode one acoustic-token frame into waveform audio, corresponding to 80~ms of speech. The total next-frame latency is computed as the sum of acoustic-token ITL and codec decoding latency.

Both codec decoders are optimized for persistent streaming inference. For the Qwen3-TTS codec, we retain the transformer KV state and causal convolution histories, cache transposed-convolution overlap, linearize streaming upsampling operations, and fuse Snake activations using native CUDA kernels with reusable FP16 buffers. For the VoiceChat-TTS codec, we use causal convolution caching, fuse the 31 codebook lookups into a single bit-exact CUDA kernel, replace channel-wise normalization sequences with native FP32 kernels, and cache the fixed inverse-STFT constants.

\setlength{\tabcolsep}{4pt}
\begin{table}[t]
\vspace{-2mm}
\caption{{Streaming latency comparison on an RTX A6000 GPU. ITL denotes the time to produce one frame of acoustic tokens at 12.5~Hz, corresponding to 80~ms of audio. Codec denotes the measured time to decode one acoustic-token frame into waveform audio.}}
\label{tab:latency_itl}
\centering
\resizebox{\columnwidth}{!}{%
\begin{tabular}{lc|cccc}
\toprule
Model 
& Conc. 
& Params 
& Acoustic ITL 
& Codec 
& Total \\
& 
& 
& (ms) $\downarrow$ 
& (ms) $\downarrow$ 
& (ms) $\downarrow$ \\
\midrule
Qwen3-TTS-12Hz~\cite{hu2026qwen3} 
& 1 
& 1.7B 
& $15.44$ 
& $4.90$ 
& $20.34$ \\
VoiceChat-TTS 
& 1 
& 977M 
& $\mathbf{7.16}$ 
& $\mathbf{2.46}$ 
& $\mathbf{9.62}$ \\
\midrule
Qwen3-TTS-12Hz~\cite{hu2026qwen3} 
& 4 
& 1.7B 
& $17.09$ 
& $4.99$ 
& $22.08$ \\
VoiceChat-TTS 
& 4 
& 977M 
& $\mathbf{12.35}$ 
& $\mathbf{2.89}$ 
& $\mathbf{15.24}$ \\
\bottomrule
\end{tabular}
}
\vspace{-2mm}
\end{table}

As shown in Table~\ref{tab:latency_itl}, VoiceChat-TTS achieves lower latency than Qwen3-TTS-12Hz under the same RTX A6000 measurement setup. At concurrency 1, VoiceChat-TTS reduces acoustic-token ITL from $15.44$~ms to $7.16$~ms, corresponding to a $2.1\times$ speedup. Its optimized codec is also faster, reducing decoding time from $4.90$~ms to $2.46$~ms. Overall, VoiceChat-TTS reduces next-frame latency from $20.34$~ms to $9.62$~ms, a $2.1\times$ improvement.

At concurrency 4, VoiceChat-TTS remains faster, reducing acoustic-token ITL from $17.09$~ms to $12.35$~ms and codec latency from $4.99$~ms to $2.89$~ms. This reduces total next-frame latency from $22.08$~ms to $15.24$~ms. All measured latencies are well below the 80~ms duration represented by one generated acoustic-token frame, indicating that both systems can run faster than real time. However, VoiceChat-TTS provides a larger latency margin for serving overheads such as batching, scheduling, network transport, playback buffering, and asynchronous codec execution, while using a smaller model.

\section{Conclusions, Limitations, and Future Work}
\label{sec:conclu}

In this paper, we introduced VoiceChat-TTS, a low-latency, continuous, and streamable text-to-speech architecture designed specifically for interactive agents. Driven directly by LLM text-token streams, VoiceChat-TTS supports explicit mid-utterance interruptions through control tokens and generates silence when no textual input is available. The model enables always-on, responsive speech generation while preserving modularity and competitive speech quality. Furthermore, it can be integrated into both duplex interaction frameworks and speech-to-speech pipelines. Our evaluations demonstrate competitive intelligibility and predicted overall speech quality relative to strong offline and streaming baselines, together with effective interruption handling and low streaming latency. VoiceChat-TTS already serves as the speech decoder in the open-source NVIDIA Nemotron VoiceChat-11B duplex S2S model, demonstrating that the proposed modular decoder can be integrated into a complete interactive speech system and support low-latency duplex interaction.

Current limitations include the absence of user-audio conditioning, which prevents the model from dynamically adapting its prosody to acoustic cues from the user, and the lack of controlled component-wise ablations. The proposed architectural changes were designed to address complementary limitations of the underlying streaming decoder, and preliminary experiments during model development indicated that their gains were cumulative. However, systematically quantifying the contribution of each component remains an important direction for future work. 
Because continuous duplex TTS is an emerging setting without standardized TTS-specific benchmarks, we also plan to develop broader evaluation protocols for long-horizon continuous generation, direct measurement of speech leakage during intended silence, and recovery across repeated interruptions.

\vfill\pagebreak
\bibliographystyle{RefStyle}

\bibliography{references.bib}

\end{document}